\documentclass[aps,prx,twocolumn,superscriptaddress]{revtex4-1}
\usepackage{graphicx}
\begin{document}

\title{Sculpting the Spin-Wave Response of Artificial Spin Ice via Microstate Selection}
\author{D. M. Arroo}
\email{d.arroo@ucl.ac.uk}
\affiliation{Blackett Laboratory, Imperial College, Prince Consort Road, London SW7 2AZ, UK}
\affiliation{London Centre for Nanotechnology and Department of Physics and Astronomy, University College London, 17-19 Gordon Street, London WC1H 0AH, UK}
\author{J. C. Gartside}
\affiliation{Blackett Laboratory, Imperial College, Prince Consort Road, London SW7 2AZ, UK}
\author{W. R. Branford}
\email{w.branford@imperial.ac.uk}
\affiliation{Blackett Laboratory, Imperial College, Prince Consort Road, London SW7 2AZ, UK}
\date{\today}

\begin{abstract}
 Artificial spin ice (ASI) systems have emerged as promising hosts for magnonic applications due to a correspondence between their magnetic configuration and spin dynamics. Though it has been demonstrated that spin-wave spectra are influenced by the ASI microstate the precise nature of this relationship has remained unclear. Recent advances in controlling the magnetic configuration of ASI make harnessing the interplay between spin dynamics and the microstate achievable. This could allow diverse applications including reconfigurable magnonic crystals and programmable microwave filters. However, extracting any novel functionality requires a full understanding of the underlying spin wave/microstate interaction.
 Here, we present a systematic analysis of how the microstate of a honeycomb ASI system affects its spin-wave spectrum through micromagnetic simulations. We find the spectrum to be highly tunable via the magnetic microstate, allowing the (de)activation of spin-wave modes and bandgap tuning via magnetic reversal of individual nano-islands. Symmetries of ASI systems and the chirality of ``monopole'' defects are found to play important roles in determining the high-frequency magnetic response.
\end{abstract}
\maketitle
\section{Introduction}

Artificial spin ice (ASI) systems consist of arrays of nanoscale ferromagnetic islands whose frustrated interactions lead to vastly degenerate low-energy states \cite{Wang2006,Nisoli2013,Heyderman2013z}. For ASI systems on a honeycomb lattice, such energy states are well characterised by the number of vertices at which islands have their magnetisations pointing either all in or all out, known as ``monopole defects'' \cite{Ladak2010,Mengotti2010,Morgan2010}. The ground state manifold consists of all magnetic configurations of the islands (``microstates'') with no monopole defect at any vertex \cite{Footnote1}.

Recently, ASI systems have emerged \cite{Gliga2013,Bhat2016c} as the potential basis of reconfigurable magnonic devices \cite{Chumak2015,Chumak2017,Krawczyk2014a} in which the high-frequency magnetic response may be altered by varying the system microstate leading to a reprogrammable magnonic band structure \cite{Iacocca2015a}. The large degeneracy of the ASI ground state could then be used to overcome the limited range of easily accessible states available in previously explored reconfigurable magnonic crystals \cite{Topp2010,Verba2012} as well as providing enhanced stability to fluctuations. Working in the other direction, measuring high-frequency eigenmode spectra could be employed as a way of ``reading out'' ASI microstates in data processing applications.

A correspondence between the high-frequency dynamics of ASI systems and their microstate would thus be beneficial for a range of applications. Though a link between microstates and high-frequency dynamics have been demonstrated for some ASI systems \cite{Gliga2015}, so far various aspects of this link remain poorly understood. In particular it is not clear from previous work under what conditions different microstates have unique spin-wave spectra.

The majority of experimental studies on the high-frequency behaviour of ASI systems to date have involved measuring ferromagnetic resonance in large arrays subject to an external swept field \cite{Bhat2016c,Jungfleisch2016,Bhat2016f, Zhou2016,Jungfleisch2017a,Bhat2018}. Since the precise magnetic microstate during such sweeps is unknown, it is unclear whether the observed changes in resonant frequencies depend only on the number of islands that have been reversed or whether the precise magnetic configuration of the array plays a role. Moreover, while the large number of magnetic microstates available in ASI systems has been posited as the source of a corresponding large number of high-frequency behaviours \cite{Jungfleisch2016}, the number of distinct regimes observed in swept-field FMR experiments so far has been modest. This may be explained by return-point memory effects in ASI systems \cite{Gilbert2015a} leading to only a small subset of the possible microstates being explored during a field sweep, but it remains to be established how many of the exponentially large number of microstates possess unique signatures in the high-frequency response.

More recent work \cite{Montoncello2018,Jungfleisch2019} has focused on isolated ASI vertices consisting of three islands at 120$^{\circ}$ to one another. These studies have clarified the contribution of individual vertices to the overall high-frequency response and highlighted how deviations from uniform magnetisation in islands can lead to mode softening, but these results also rely on swept fields and since they concern individual vertices they cannot explore the rich microstate space and long-range interactions of extended ASI arrays.

Here, the link between the microstate of an ASI system and its high-frequency dynamics is analysed in detail via micromagnetic simulations, with a view to isolating the role of the ASI microstate in the spin-wave response from effects of external magnetic fields. The results are relevant to assessing how ASI systems may be incorporated into the ever growing range of reconfigurable magnonic device concepts for low-power information processing \cite{Krawczyk2014a}.

\section{Simulation Methodology}

Spin-wave eigenmode spectra were generated for honeycomb ASI systems with MuMax3 \cite{Vansteenkiste2014} by applying an out-of-plane field pulse to and sampling the evolution of the system's magnetisation every 20 ps over the subsequent 5 ns. The pulse applied was uniform in space (so that excited modes correspond to $\vec{k}=0$ uniform precession modes) and varied as a sinc function in time with a maximum amplitude of 20 mT, so that frequencies up to 25 GHz were excited equally. By applying a Fourier transform to the magnetisation of each micromagnetic cell, the intensity of precession can the be mapped out as a function of both frequency and position \cite{Dvornik2013,Baker2016}.

\begin{figure}
	\centering
	\includegraphics[width=1.0\linewidth]{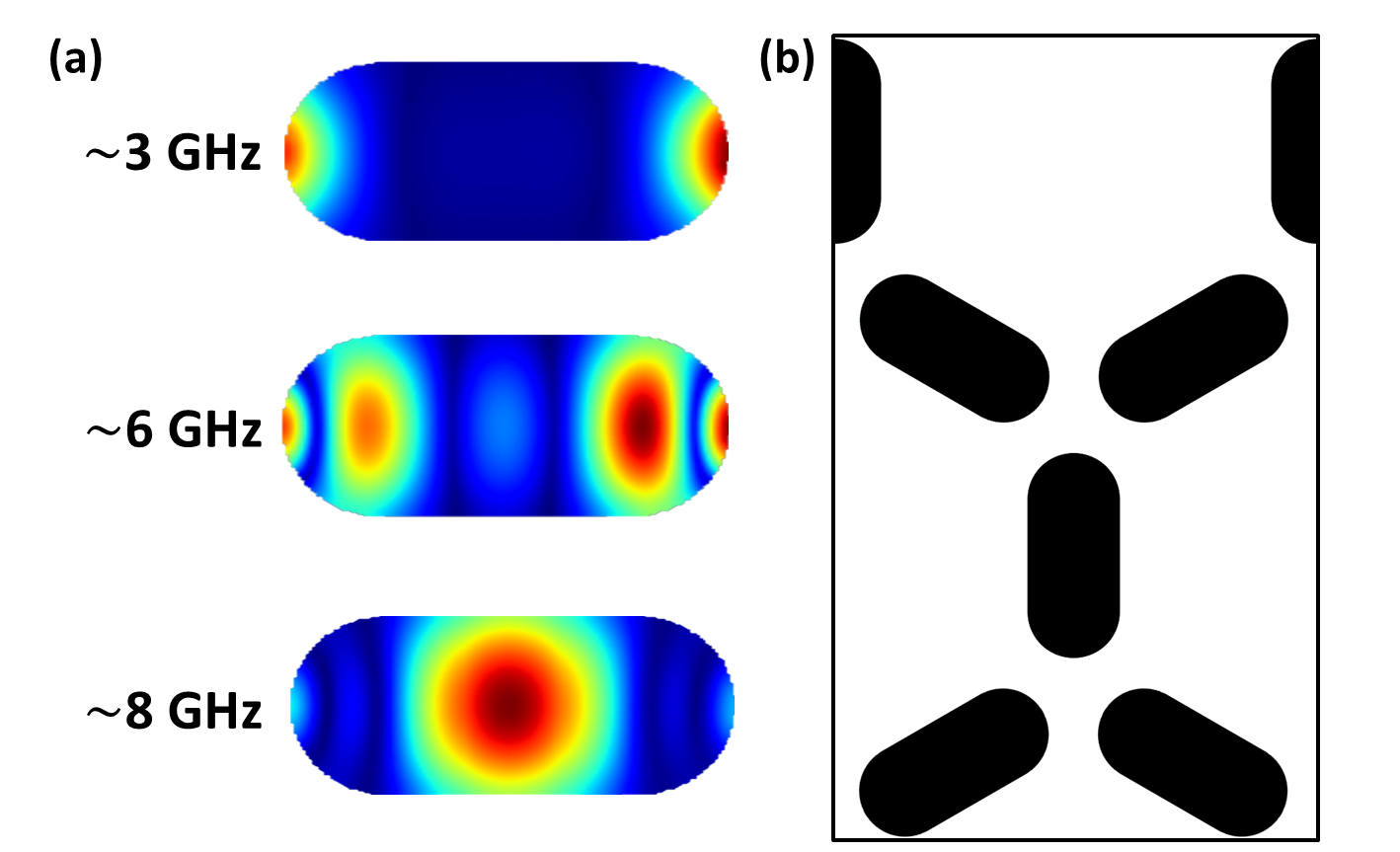}
	\caption{(a) Power distributions for the three eigenmodes observed for an isolated ferromagnetic island. The mode around 8 GHz is localised in the bulk of the island and has the largest amplitude, while the two lower frequency modes are localised at the edges. (b) A six-island artificial spin ice unit cell simulated with periodic boundary conditions, allowing $2^{6}$ configurations with up to four monopole defects per unit cell to be explored.}
	\label{fig:SingleBar-Modes}
\end{figure}

Simulations were performed for 130 nm $\times$ 290 nm $\times$ 20 nm islands with micromagnetic parameters matching those of permalloy, with $M_{\text{S}}= 800$ kA/m, $A_{\text{ex}}=13$ pJ/m and $\alpha=0.006$. The basic form of the eigenmode spectra was found to derive from that of the single-island spectrum, exhibiting three modes: two localised at the island edges and one in the bulk (Fig. \ref{fig:SingleBar-Modes}a). In an ASI array, magnetostatic interactions caused these modes to shift in frequency and split into multiplets.

To gain an overview of the effect of the magnetic configuration on the high-frequency dynamics of an extended honeycomb ASI array, the system was reduced to a plaquette of six islands with four vertices (Fig. \ref{fig:SingleBar-Modes}b) and simulated with periodic boundary conditions. Because the magnetisation of the ferromagnetic islands in ASI systems is strongly constrained by shape anisotropy to lie along the island-axis, the islands are commonly treated as macrospins \cite{Nisoli2013}, giving a total of $2^{6}=64$ configurations for a six-island system. Spectra were generated at remanence for the entire configuration space, allowing the effect on the eigenmode spectrum of a range of different configurations and monopole densities to be probed.

\section{Results and Discussion}

\begin{figure}
	\centering
	\includegraphics[width=1.0\linewidth]{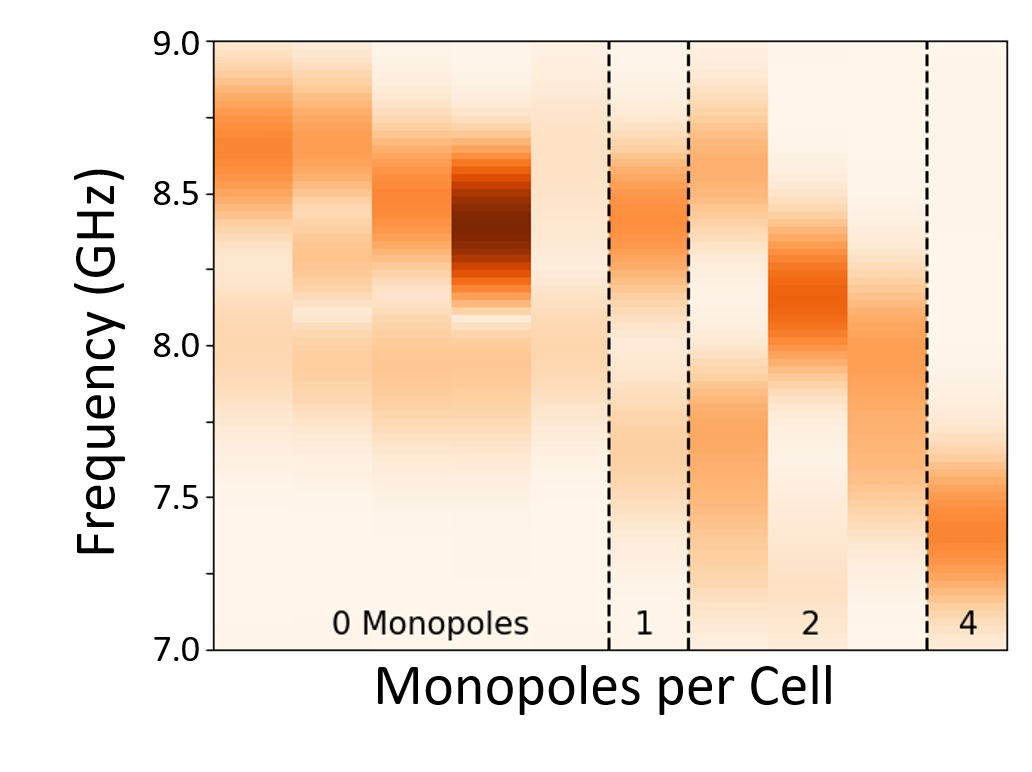}
	\caption{Spectra around the ~8 GHz bulk mode for the 10 microstates that are distinguishable in the macrospin picture for the six-island unit cell shown in Fig. \ref{fig:SingleBar-Modes}b, arranged in order of monopole density. The remaining 54 available microstates may be mapped onto these ten microstates via rotations and global spin reversals, and reproduce their corresponding spectra.}
	\label{fig:allModes}
\end{figure}

Within a macrospin picture, the number of microstates with distinguishable mode spectra is significantly reduced by the fact that many configurations may be mapped onto one another via global spin reversals and rotations. The dynamics of ferromagnetic systems is described by the Landau-Lifshitz equation 

\begin{equation}
\frac{\partial\mathbf{M}}{\partial t}=\gamma\mu_{0}\mathbf{M}\times\mathbf{H}_{\text{eff}}
\end{equation}
which has no explicit dependence on direction and is invariant under time-reversal (i.e., a global spin reversal), so that microstates separated only by these transformations have identical spectra. The time-reversal symmetry is in fact broken when damping is included, but in a steady state the damping is precisely cancelled out by the driving ac-field so that it can be neglected.

For a finite honeycomb system with $n$ macrospins and full hexagonal symmetry the number of spectrally distinguishable microstates is thus reduced from $2^{n}$ to $2^{n/2}$, putting a limit on the versatility of ASI systems as reconfigurable magnonic components. Here a reduced six-island system (Fig. \ref{fig:SingleBar-Modes}b) is studied with periodic boundary conditions, so that translational symmetry and the broken hexagonal symmetry of the rectangular plaquette that was used leave 10 configurations with distinct spectra. These 10 configurations may be meaningfully sorted by the number of monopole defects per cell, since the total monopole number is invariant under both rotations and global spin reversals.

The ASI spectra generated showed bands corresponding to the three modes observed in an isolated ferromagnetic island (the band corresponding to the bulk mode around 8 GHz is shown in Fig. \ref{fig:allModes}), with inter-island interactions causing causing peaks to shift in frequency or split into multiple peaks depending on the microstate. It was found that for the bulk mode around 8 GHz the macrospin picture indeed accounts for 10 distinct spectra, whereas for the edge modes localised curling of the magnetisation leads to a departure from the macrospin description which breaks the rotational symmetry of the system and gives an additional degree of freedom \cite{Rougemaille2013}. 

\subsection{Bulk Modes}

Focusing on the 8 GHz bulk mode (Fig. \ref{fig:allModes}), changes in configuration can lead to the splitting of modes as well as shifts in the peak centre of more than 1 GHz. The maximum shift is thus comparable to previously reported shifts \cite{Topp2010,Verba2012} in reconfigurable magnonic crystals, though honeycomb ASI offers the additional benefit of a large number of intermediate configurations that allow the peak position to be smoothly tuned from one position to another. A general downward trend in the position of peak centres as the monopole number increases may be explained by the reduced effective field, $\mathbf{H}_{\text{eff}}$, felt by islands as the opposing stray field from neighbouring islands increases. 

\begin{figure}
	\centering
	\includegraphics[width=1.0\linewidth]{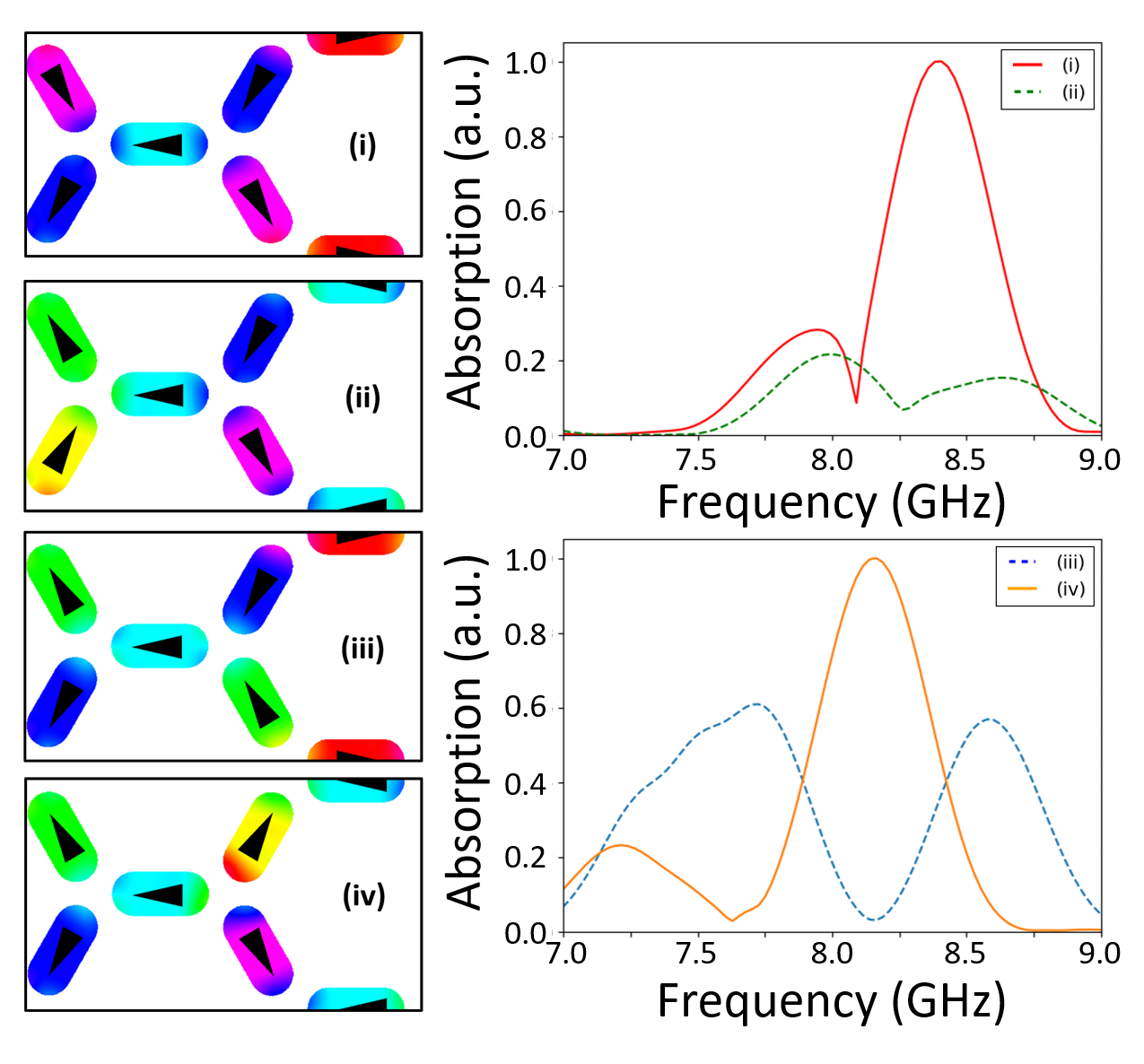}
	\caption{Moving between two zero-monopole configurations (i, ii) or two 2-monopole configurations (iii, iv) leads to significant enhancement/suppression of the 8 GHz bulk mode, suggesting applications as ``ON'' and ``OFF'' states. The enhancement factor depends on the strength of the inter-island coupling}
	\label{fig:on-off}
\end{figure}

In addition to shifting and splitting modes, changing the magnetic microstate allows the amplitude of the modes to be enhanced or suppressed (Fig. \ref{fig:on-off}) suggesting applications as microwave filters or ``ON/OFF'' element in magnonic circuits. Such ``ON'' and ``OFF'' microstate-pairs exist in both the zero- and two-monopole states and allow enhancements by a factor of up to 30 without any net change in the total system energy, comparable to recent results in a 1D reconfigurable magnonic device \cite{Haldar2016}. 

That it is possible to distinguish between states with the same monopole density is testament to the exquisite sensitivity of the FMR response to the internal field distribution. The distinct spectra observed for the five different monopole-free microstates may be explained in terms of the vector sum of the stray fields felt by each island from its four nearest neighbours.
 
For each vertex obeying the ice rule constraint (two-in/one-out or one-in/two-out), there are two islands whose magnetisation is oriented in the majority spin direction (i.e. "in" for a two-in/one-out vertex) and one with magnetisation in the minority spin direction ("out"). Whether an island is oriented in the majority or minority direction for a particular vertex determines the stray field it feels from the other two islands at the vertex: minority islands experience a stray field parallel (``='') to their magnetisation direction whereas the stray field for majority islands in one of the directions perpendicular (``$\uparrow$'' or $\downarrow$'') to their magnetisation \cite{SuppInfo1}. Since each island subtends two vertices this means that each island experiences one of four possible combinations of perpendicular and parallel fields, which in turn may combine in one of five ways to to form a monopole-free microstate on the unit cell studied here. The combinations of different fields felt by the six island determine the fine structure of the five 0MP spectra in Fig. \ref{fig:allModes} - in particular, the bulk mode may be resolved into two or three sub-modes where the number of sub-modes depends on the number of distinct fields felt by the six islands.

Such differences between microstates with the same monopole density indicate that even in the part of the island where inter-island interactions are sufficiently weak that the macrospin approximation holds, they are nevertheless involved in modifying the high-frequency response. The role of the stray fields from neighbours is further supported by the fact that the enhancement/suppression factor between two states could be made larger by increasing the saturation magnetisation, $M_{\text{S}}$, so that inter-island interactions were stronger. It was possible to increase the enhancement/suppression ratio by a further factor of four by increasing $M_{\text{S}}$ to 1200 kA/m. Beyond this the simulated islands were no longer stable as single ferromagnetic domains, but exploring ferromagnetic materials with high saturation magnetisations such as cobalt and tuning the gap between neighbouring islands is a viable route to maximising the enhancement/suppression ratio.

\subsection{Edge Modes}

For the two modes at around 3 GHz and 6 GHz which are localised at the edges of islands, there is an additional degree of freedom in the way that the magnetisation of the nano-islands curls up at vertices in order to minimise the magnetostatic self-interaction and interactions with neighbouring islands. In this case the macrospin approximation is no longer strictly valid \cite{Rougemaille2013,Gliga2015} and two configurations which map onto one another as macrospins may have different spectra if their island edges curve differently at vertices.

At each vertex three edges come together which may either curve (C) or remain straight as for individual islands (S), leading to four possible vertex types: SSS, SSC, CCS and CCC \cite{SuppInfo2}. The vertex types that include curved edges may then be further characterised by whether or not all their edges curve in the same direction and if so by the sense of their chirality. This gives eleven vertex types in total and in principle $11^{4}$ different vertex configurations on the periodic plaquette used in this study, offering substantial scope for tuning the high frequency response of honeycomb ASI by controlling the edge curling.

Not all of these vertex types are stable in the absence of an external field and in simulations carried out at remanence monopoles were only seen in the two chiral CCC states, while 2-in/1-out and 1-in/2-out vertices were only observed in achiral CCS states with the two islands with the majority spin direction showing curling in opposite directions. Unstable vertex types may nevertheless be observed in ASI systems if they are stabilised using an external magnetic field \cite{Rougemaille2013}.

The precise extent to which the magnetisation curls at island-edges depends on the competition between magnetostatic inter-island and self-interactions. These interactions may be tuned via the inter-island separation and the island thickness, respectively, with the edge curling vanishing when simulations were repeated with the island thickness set below 3 nm or the edge-to-edge separation between islands increased beyond 50 nm (i.e. where the dipolar field from neighbouring edges becomes small compared to the magnetisation of permalloy).

\begin{figure}
	\centering
	\includegraphics[width=1.0\linewidth]{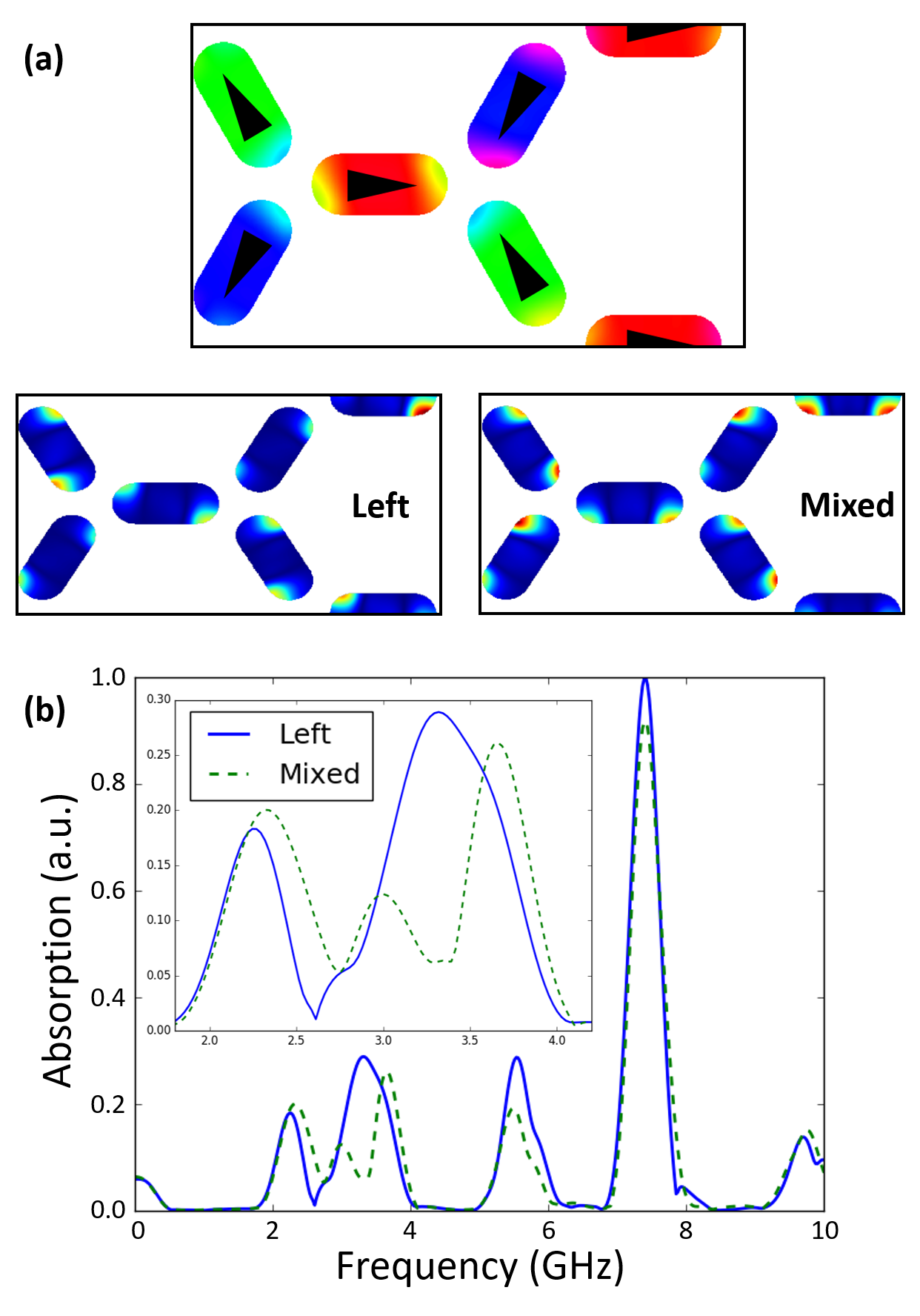}
	\caption{(a) The 4-monopole microstate (top left) was prepared with two different edge curling profiles, the effect of which can be seen in the distorted spatial profile of the 3 GHz mode (top right, bottom). (b) Eigenmode spectra for the 4-monopole microstate with the left-handed and mixed edge curling profiles. The 3 GHz mode (inset) is particularly affected by moving between the different profiles.}
	\label{fig:MonoSpectrum}
\end{figure}

To determine whether the edge-localised modes are sensitive to this edge curling the 4-monopole microstate was prepared with two different magnetisation profiles: one with only ``left-handed'' vertices and one in which right- and left-handed vertices alternate (Fig. \ref{fig:MonoSpectrum}a). The lower-frequency edge mode was found to indeed be sensitive to edge curling while the 6 GHz edge mode was not observed to vary significantly as a result of the different magnetisation profiles, the 8 GHz bulk mode was insensitive to edge curling as expected (Fig. \ref{fig:MonoSpectrum}b).

That the higher-frequency edge modes did not respond to the different edge curling profiles as much as the 3 GHz modes is unexpected as their spatial profiles show a similar chiral distortion to that seen for the 3 GHz mode (Fig. \ref{fig:MonoSpectrum}a), but is likely to be because less of the power associated with the higher-frequency edge modes is localised right along the island boundaries where curling is strongest (Fig. \ref{fig:SingleBar-Modes}a).

The 3 GHz edge mode was substantially altered by moving between the different edge curling profiles. It is notable that the ``mixed'' profile spectrum is not simply a linear combination of the right- and left-handed profiles, attesting to the importance of edge curling in inter-island interactions and in accessing the full range of high-frequency behaviours available in ASI systems \cite{Li2016d}. We note that in addition to an intrinsic effect on the spectrum from the magnetic configuration, real ASI systems will be display additional extrinsic differences in the spectra of states with varying edge curling profiles due to roughness introduced in the fabrication that breaks symmetries. The edge curling thus creates distinct FMR fingerprints which may form the basis of a readout method with which to characterise the chirality of monopoles in ASI systems.

\subsection{Effect of an External Magnetic Field}

It was found that at remanence monopoles are only stable in chiral CCC states while unexcited vertices are only stable in an achiral CCS state, limiting the edge curling profiles available for controlling high-frequency dynamics. However, the remaining vertex states may still be accessed if they are stabilised by an external magnetic field, with the additional consequence that the external field will break the rotational and time-reversal symmetries and allow configurations that map onto each other to be distinguished from each other spectrally. 

The effect of an external magnetic field (applied parallel to the long axis of the honeycomb cell) on the eigenmode spectra of an ASI system with a reduced lattice parameter were simulated for a polarised state and the 4-monopole state to demonstrate how the field and microstate interact to determine the high-frequency response. The spectra for these systems under a range of external fields from -100 mT to +100 mT are shown in Fig. \ref{fig:fig3}, where for each field the microstate was re-initialised and allowed to relax so that the spectra are indicative of the expected response if the system is initialised and immediately subjected to a given field with no further field history.

The evolution of individual modes of the polarised configuration are generally well-described by the Kittel formula for ferromagnetic resonance:
\begin{equation}\label{key}
f(H)\propto\sqrt{(M_{\textrm{s}}+H_{\textrm{k}})(M_{\textrm{s}}+H_{\textrm{k}}+H)}
\label{eq:kittel}
\end{equation}
where $M_{\textrm{s}}$ denotes the saturation magnetisation, $H_{\textrm{k}}$ the effective field due to magnetic anisotropy and $H$ is the externally applied field at which resonance occurs. Since the effective anisotropy field is increased by reducing the lateral dimensions of each island, the frequencies of the modes present at zero-field in Fig. \ref{fig:fig3} are shifted upwards with respect to their positions in Fig. \ref{fig:SingleBar-Modes}. The degeneracy between the modes of the diagonal and horizontal islands is further broken due to the broken rotational symmetry, leading to a fully separate bulk mode for the diagonal islands in which $H_{\textrm{k}}$ does not align with the external field $H$. Rotational symmetry breaking may thus be used to introduce additional eigenmodes.

\begin{figure}
	\centering
	\includegraphics[width=1.0\linewidth]{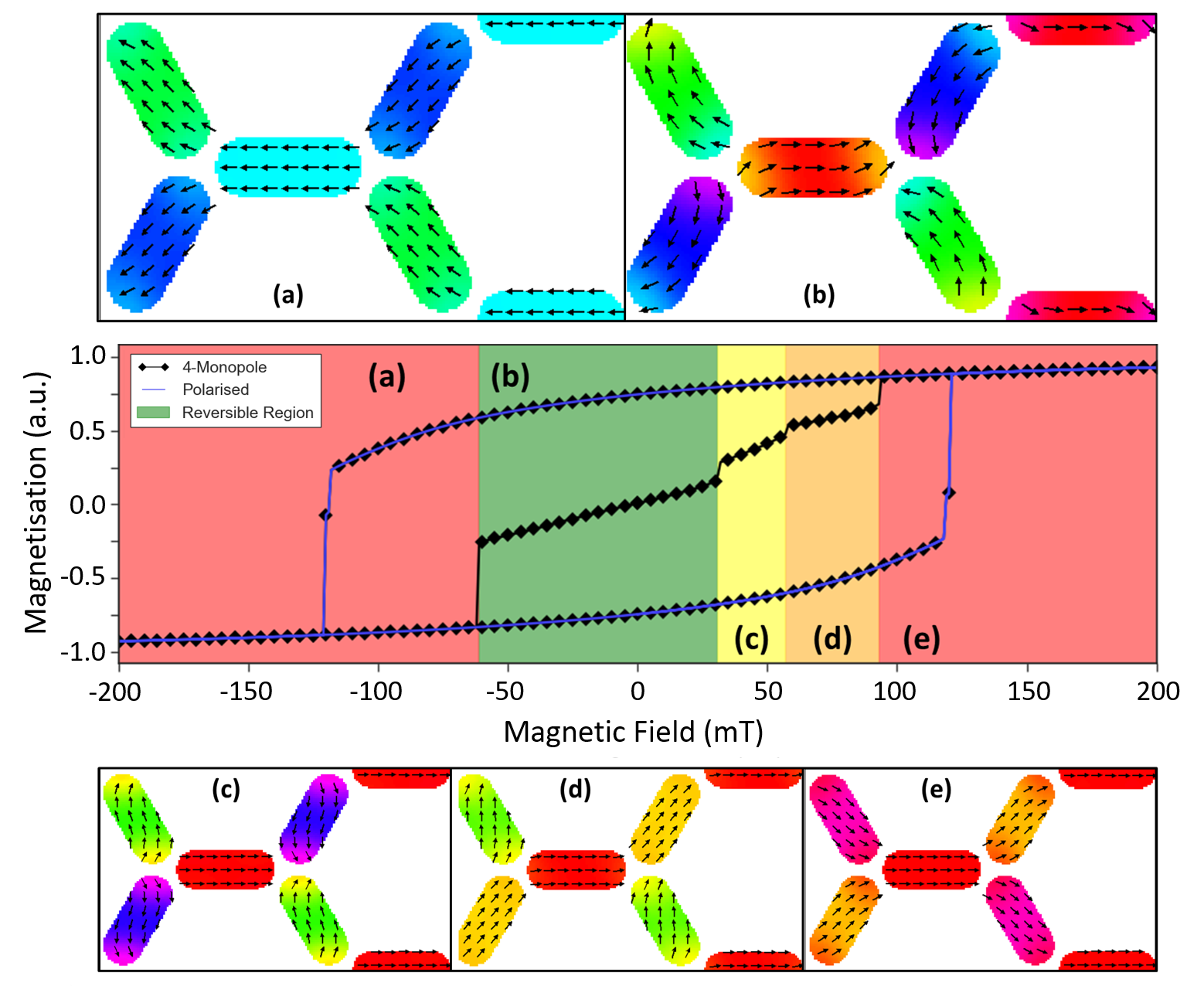}
	\caption{Magnetisation curves for a honeycomb ASI system initialised at zero field in a polarised and 4-monopole configuration and allowed to relax in a magnetic field applied along the long axis of the unit cell. The field intervals over which changes in the magnetisation of both configurations are reversible are shaded, showing five clear regimes (a) left-polarised; (b) initial 4-monopole state (with a mixed edge curling profile); (c) 4-monopole state (with field-stabilised achiral monopoles); (d) ice-rule state; (e) right-polarised. In regions (a) and (e) the two curves coincide since the 4-monopole state is fully saturated.}
	\label{fig:mhplusconfigs-labeled}
\end{figure}

\begin{figure}
	\centering
	\includegraphics[width=1.0\linewidth]{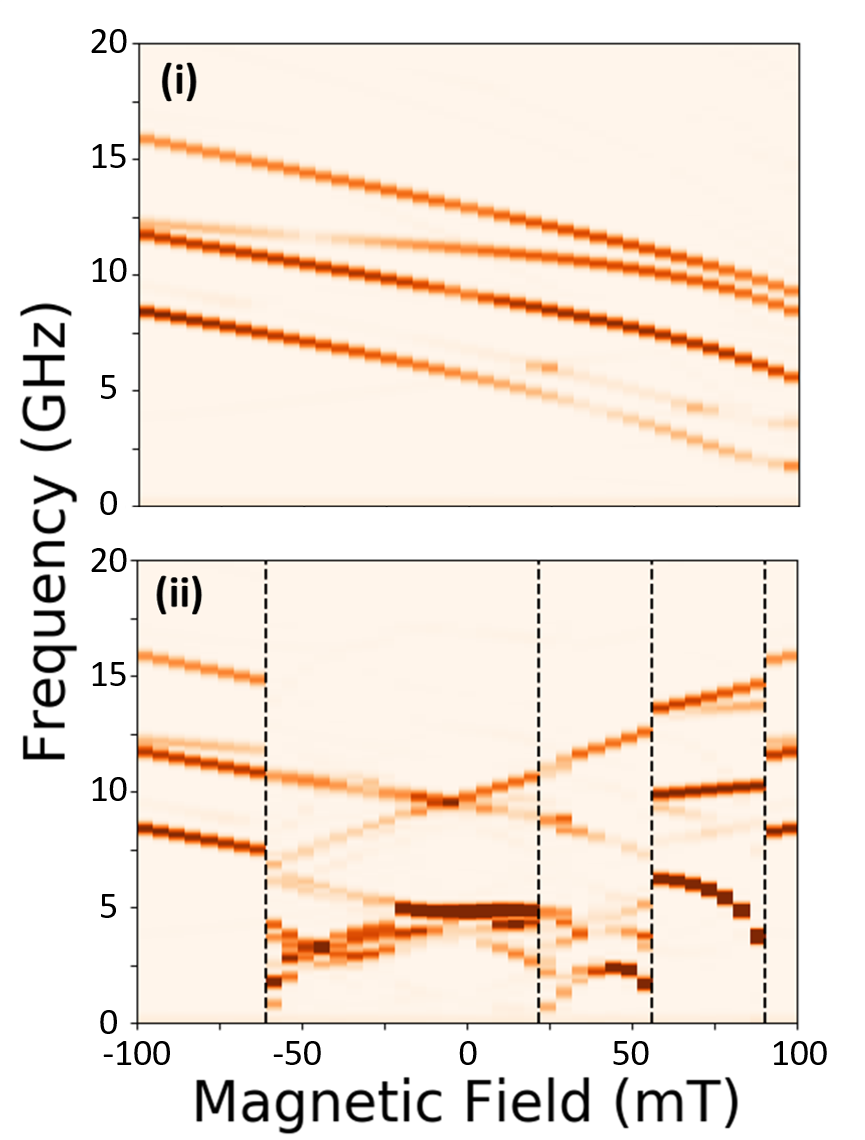}
	\caption{Eigenmode spectra for systems initialised in the (i) polarised and (ii) 4-monopole configurations (shown in Fig. \ref{fig:mhplusconfigs-labeled}a and \ref{fig:mhplusconfigs-labeled}b, respectively) as a function of an externally applied field. The spectrum initialised in the 4-monopole state shows the same five regimes as observed in Fig. \ref{fig:mhplusconfigs-labeled}.}
	\label{fig:fig3}
\end{figure}

For the 4-monopole state, the evolution of the eigenmode spectrum as an external field is applied is more complex as the edge curling and eventually full magnetisation of islands reverses to align with the applied field (Fig.  \ref{fig:mhplusconfigs-labeled}). At around 30 mT in the positive $x$-direction, achiral monopoles form and are stable for as long as the field is applied (Fig. \ref{fig:mhplusconfigs-labeled}). As was observed at remanence only the lowest-frequency edge mode, which is here redshifted by over 1 GHz, is sensitive to a change in the edge curling profile (Fig. \ref{fig:fig3}). At just above 50 mT, two diagonal islands irreversibly switch their magnetisation direction and a small jump in the total magnetisation triggers a remarkable change in the eigenmode spectrum, with the middle mode shifting by almost 3 GHz. The system saturates at 90 mT (or 60 mT applied in the negative $x$-direction) at which point its mode spectrum reproduces that of the polarised system.

The application of external fields thus offers an additional means of controlling the high-frequency response of ASI systems that complements microstate selection. Since field-stabilised vertex curling profiles will relax back into their initial states when the external field is removed, it is possible to achieve discrete shifts in the high-frequency response without leaving the reversible region beyond which the microstate is lost. The microstate may therefore be selected to broadly determine the high-frequency characteristics using an external field applied to produce additional modes by breaking symmetry. External fields may further modulate this response either in a continuous fashion (following Eq. \ref{eq:kittel}) or in reversible discrete jumps mediated by changes in the vertex curling profile.

\section{Conclusion}

In summary, the high-frequency response of honeycomb ASI in different magnetic configurations has been characterised via micromagnetic simulations. For modes localised in the bulk of ferromagnetic islands, the dependence on the magnetic microstate of the system is well-described within a macrospin approximation. A consequence of this is that the number of microstates with distinguishable spectra at remanence is lower than the number of microstates itself, since the eigenmode spectra are invariant under rotations and global spin-reversals. This redundancy in the microstates activating a particular spectrum offers increased flexibility in the control of high-frequency dynamics, allowing the most convenient microstate to be selected from a range of isospectral alternatives where microstate selection is constrained. The precise degree of redundancy may be altered by exploring ASI systems with broken rotational symmetry \cite{Morrison2013,Chern2016,Dion2019}.

Within the distinguishable microstate space significant control of the spin-wave spectrum was achieved by moving between different microstates. The bulk mode may be shifted by as much as 1 GHz, split into discrete modes or switched off by moving between the 10 non-redundant microstates available on the six-island plaquette studied here. Stronger mode enhancement/suppression is achievable in systems where the inter-island interaction strength is increased. Increasing the number of nano-islands introduces further degrees of configurational freedom, offering even greater scope for sculpting the dynamic response which scales extensively with system size. The large number of near-degenerate configurations in ASI systems mean that the spin-wave spectrum may be smoothly tuned - in particular increasing the monopole defect density gradually lowers the frequency of the bulk mode as opposing magnetostatic fields reduce the effective field felt by the system, with the ``resolution'' of tuning improving with system size. This demonstrates the benefits of  large, fully-accessible microstate space and overcomes a key limitation of previously explored reconfigurable magnonic crystal concepts \cite{Topp2010,Verba2012}, which support only a limited number of stable configurations.

Modes localised at island edges were found to be sensitive to curling in the magnetisation neglected in the macrospin picture. This edge curling provides an additional degree of freedom with which to control the high-frequency response, with the application of external fields stabilising different types of vertex curling. External fields also break rotational and time-reversal symmetry, introducing additional modes and allowing further tuning of high-frequency dynamics by breaking degeneracies resulting from the symmetries of a macrospin system.

A combination of global external fields and local nano-island stray fields determined by the ASI microstate thus allows substantial control over the high-frequency response of ASI systems. The significance of this is heightened by recently developed techniques for state-control \cite{Burn2016} and island-specific magnetic writing \cite{Gartside2017d,Wang2016,Vavassori2019}, bringing fine spectral control within experimental reach by providing a method of reliably moving between microstates. The macroscopic degeneracy of ASI systems ensures a colossal range of stable configurations whose varying properties may be selected to precision-tune the response of prospective devices ranging from programmable microwave filters to reconfigurable magnonic components. The lattice symmetry and symmetry-breaking effects of inter-island interactions provides a broad range of parameters for the rational design of versatile, highly tunable spectrum-selection devices inviting a host of applications from microwave filters to reconfigurable functional magnonic elements for low-power data processing.

\begin{acknowledgments}
This work was supported by the Engineering and Physical Sciences Research Council (grant number EP/G004765/1) and the Leverhulme Trust (grant number RPG 2012-692) to WRB. 

The authors would further like to thank Professor Lesley F. Cohen for illuminating discussions and the Imperial College Research Computing Service (DOI: 10.14469/hpc/2232) for access to their HPC facilities. 
\end{acknowledgments}

\bibliographystyle{unsrt}

\begin{thebibliography}{10}
\bibitem{Footnote1}
	It should be noted that in some previous studies, the term ``monopole'' has been used to describe excitations with respect to a reference background state, so that in honeycomb ASI a two-in/one-out vertex would be counted as a monopole if the same vertex had a one-in/two-out in the reference state. In this work, we reserve the term ``monopole'' for vertices with an all-in or all-out configuration, so that they act as a convenient way to characterise the total energy of the system with no need to specify a reference state.
\end{thebibliography}

\end{document}